\documentclass[sigconf]{acmart}

\renewcommand\footnotetextcopyrightpermission[1]{} % removes footnote with conference information in first column
\usepackage{array}
\newcolumntype{L}[1]{>{\raggedright\let\newline\\\arraybackslash\hspace{0pt}}m{#1}}
\newcolumntype{C}[1]{>{\centering\let\newline\\\arraybackslash\hspace{0pt}}m{#1}}
\newcolumntype{R}[1]{>{\raggedleft\let\newline\\\arraybackslash\hspace{0pt}}m{#1}}
\usepackage{tabularx}

\usepackage{booktabs} % For formal tables
\usepackage{ragged2e} % for tighter hyphenation
\setcopyright{rightsretained}
\acmConference[WWW'19]{WWW}{2019}{San Francisco, USA}
\acmYear{2019}
\copyrightyear{2019}

\usepackage{url}
\usepackage{hyperref}
\usepackage{enumitem}
\usepackage{flushend}
\usepackage{textcomp}
\usepackage{fixltx2e}
\usepackage{siunitx}
\begin{document}
\title{Examining the Impact of Algorithm Awareness on Wikidata's Recommender System Recoin}

\author{Jesse Josua Benjamin}
\affiliation{Human-Centered Computing \\ Freie Universit\"at Berlin
}
\email{jesse.benjamin@fu-berlin.de}

\author{Claudia M\"uller-Birn}
\affiliation{Human-Centered Computing \\ Freie Universit\"at Berlin
}
\email{clmb@inf.fu-berlin.de}

\author{Simon Razniewski}
\affiliation{Max Planck Institute for Informatics}
\email{srazniew@mpi-inf.mpg.de}

\begin{abstract}
The global infrastructure of the Web, designed as an open and transparent system, has a significant impact on our society. However, algorithmic systems of corporate entities that neglect those principles increasingly populated the Web. Typical representatives of these algorithmic systems are recommender systems that influence our society both on a scale of global politics and during mundane shopping decisions. Recently, such recommender systems have come under critique for how they may strengthen existing or even generate new kinds of biases. To this end, designers and engineers are increasingly urged to make the functioning and purpose of recommender systems more transparent. Our research relates to the discourse of algorithm awareness, that reconsiders the role of algorithm visibility in interface design. We conducted online experiments with 105 participants using MTurk for the recommender system Recoin, a gadget for Wikidata. In these experiments, we presented users with one of a set of three different designs of Recoin's user interface, each of them exhibiting a varying degree of explainability and interactivity. Our findings include a positive correlation between comprehension of and trust in an algorithmic system in our interactive redesign. However, our results are not conclusive yet, and suggest that the measures of comprehension, fairness, accuracy and trust are not yet exhaustive for the empirical study of algorithm awareness. Our qualitative insights provide a first indication for further measures. Our study participants, for example, were less concerned with the details of understanding an algorithmic calculation than with who or what is judging the result of the algorithm. 
\end{abstract}


% <concept_id>10003120.10003121</concept_id>
% <concept_desc>Human-centered computing~Human computer interaction (HCI)</concept_desc>
% <concept_id>10003120.10003121.10011748</concept_id>
% <concept_desc>Human-centered computing~Empirical studies in HCI</concept_desc>
% <concept_significance>300</concept_significance>
% <concept_id>10002951.10003317.10003347.10003350</concept_id>
% <concept_desc>Information systems~Recommender systems</concept_desc>
% <concept_significance>300</concept_significance>
% <concept_id>10003033.10003106.10003114.10003116</concept_id>
% <concept_desc>Networks~World Wide Web (network structure)</concept_desc>
% <concept_significance>300</concept_significance>

% \ccsdesc[500]{Human-centered computing~Human computer interaction (HCI)}
% \ccsdesc[300]{Human-centered computing~Empirical studies in HCI}
% % \ccsdesc[300]{Human-centered computing~Computer supported cooperative work}
% \ccsdesc[300]{Information systems~Recommender systems}

\maketitle

\section{Motivation and Background}
After three decades of continuous growth~\cite{woodall_measuring_2017}, the Web has become an integral part of our society. It is designed as an open and transparent system, but more recently algorithmic systems that neglect these principles populate the Web. Exemplary of these systems are recommender systems, which have different impacts ranging from societal discourses (e.g., the British EU referendum or the U.S. presidential election of 2016) to profane details of everyday life, such as when choosing a product or service, a place to spend the holidays, or consuming personalized entertainment products. On the whole range of their usage, recommender systems are often interpreted as algorithmic decision support systems that frequently include discussions on bias (e.g.,~\cite{oneil_weapons_2016}). One of these frequently raised issues is that of \emph{algorithmic bias}, where specific groups of people based on gender, ethnicity, class or ideology are systematically discriminated by algorithmic decisions~\cite{baeza-yates_bias_2018}. These discussions indicate a growing discomfort with the algorithmic advances both used in and facilitated by the Web. 

To this end, ongoing design discourses urge engineers to consider explaining the presence and function of algorithms to end users~\cite{kuang_next_2017}. Lawmakers, too, are increasingly called upon to respond to issues such as algorithmic bias. Exemplary of the latter is the General Data Protection Regulation (GDPR) by the European Union, which seeks to curtail violations of data privacy and unregulated profitisation from user data. Selbst and Powles go so far as to say that the GDPR effectively guarantees a ``right to explanation'' of algorithmic processes to end users~\cite{selbst_meaningful_2018}. 

This right to explanation is understood as an explicit challenge to the discipline of Human-Computer Interaction (HCI); to be met with concrete means of representing and explaining algorithmic processes. In HCI, this challenge aligns with the discourse of \textit{algorithm awareness}. Hamilton and colleagues define algorithm awareness as the extent to which a user is aware of the existence and functioning of algorithms in a specific context of use~\cite{hamilton_path_2014}. 

%However, while algorithm awareness\footnote{The concept of Algorithm Awareness belongs to a broader discussion that comprises Fairness, Accountability and Transparency (FAT), Explainable Artificial Intelligence (XAI), and Critical Data Studies (CDS).} has become the focus of contemporary studies, existing approaches face a number of limitations.
%Previous studies on algorithm awareness employ methods such as reverse engineering~\cite{eslami_understanding_2017} mainly because major web-based systems (e.g., Facebook) hide the functioning of their algorithms. Other studies rely on concepts from justice theory, such as Kizilcec~\cite{kizilcec_how_2016}.}\clmb{Jesse, Why is this a limitation? - Please explain in more detail.
However, the scope of the term algorithm awareness is not yet clearly defined, partially a result of the lack of experimental results associated with the discourse. As a consequence, it is unresolved whether algorithm awareness is the result of unearthing new methods of interaction, novel forms of representation, finding means of explaining algorithmic processes, or all of these aspects taken together. Similarly, its methodological perspective is vaguely defined. Are algorithm-aware designs, for example, a result of a critical technical practice~\cite{sengers_reflective_2005}, or are they a new form of human-centered design? If algorithm awareness as a principle is to contribute to an understanding of web-based algorithmic systems today (and tomorrow), these methodological shortcomings need to be addressed.

In our research, we focus on one specific aspect from the discourse of algorithm awareness, the aspect of algorithm representation. We first discuss related work from the areas of HCI, Computer-Supported Collaborative Work (CSCW), and Science and Technology Studies (STS). We argue that algorithm awareness should be understood within the context of human-technology relations since algorithmic systems increasingly impact how we see the world. We then introduce a use case which allows for studying different representations for algorithm awareness because of its open design. The use case is situated in the peer production system Wikidata, in which the completeness recommender Recoin is being used by editors to receive recommendations for their next edits. As opposed to commercial web-based systems, the design principles of Wikidata give as access to all necessary information available regarding Recoin's engineering and usage. We can, thus, reflect on the various decisions made during Recoin's development and can suggest different modes of representing the algorithmic system by considering the dimensions of explainability and interactivity. %Additionally, by providing a use case from peer production systems, we can argue more substantially about the role of algorithm awareness with regards to socio-technical dimensions. 

Our research makes the following contributions: We provide experimental results to be used in the continuing development of the discourse on algorithm awareness. This concerns insights on design measures, namely textual explanation of algorithmic logic and interactive affordances, respectively. Our results suggest that the provision of more interactive means for exploring the mechanism of the algorithm for users has significant potential, but that more research is needed in this area. As for conducting experiments in this context, we provide first methodological insights, which suggest that measures of comprehension, fairness, accuracy and trustworthiness employed in the field are not yet exhaustive for the concerns of algorithm awareness. Our qualitative insights provide a first indication for further measures. The study participants, for example, were less concerned with the details of understanding an algorithmic calculation than with who or what is judging the result of the algorithm. 

%\begin{enumerate}
%\item \textcolor{red}{What does awareness mean in terms of algorithmic processes? - needs discussion}
%\end{enumerate}
%In practical terms, we consider:
% \begin{enumerate} 
% \item What effect does the recommender Recoin have on contributions to Wikidata?
% \item To what extent differ contributions to Wikidata depending on the type of algorithm representation in Recoin?
%  \item How do different types of algorithm representation affect human understanding of the functioning of the underlying algorithmic system?
%\end{enumerate}

In the following section, we discuss existing research from three perspective. First, we discuss the role of automation in peer production communities which leads to an increased usage of algorithmic systems in this context. Second, we review existing approaches that attempt to make these algorithmic systems transparent. Third, we combine these insights to argue for the urgency of researching algorithm awareness. This theoretical section is followed by a detailed introduction to Recoin, our use case, including its technical design as well as how it connects specifically to the topic at hand. Subsequently, we showcase our experiment, in which we detail the setup, design, results and analyses involved in our experimental study. Then, we proceed to discuss our insights. Finally, we conclude by outlining future work, in which we seek to undertake qualitative studies into how the evaluated modes of representation affect the relations between humans and algorithmic systems. 

%******************************************************************************************************************************************************
\section{Related Work}
\label{sec:relatedwork}
\paragraph{\textbf{Automation in Peer Production Communities}}
In contrast to the predominant commercial platforms on the Web, peer production communities, such as Wikipedia, OpenStreetMap, or Linux, provide a valuable alternative for people to share their ideas, experiences, and their collaboratively created knowledge openly~\cite{Benkler:2002vn,Benkler:2006pi}. In these communities automation is an integral component in order to handle especially "mindless, boring and often reoccurring tasks" \cite{Mueller-Birn:2013uq}. In Wikipedia, for example, various forms of algorithmic support exist; recommender systems, such as the SuggestBot help people to find suitable tasks~\cite{Cosley:2007fk}, neural networks, such as employed by ClueBots help to revert obvious vandalism~\cite{carter2008cluebot}, or semi-automated user interfaces, such as Snuggle help editors to socialize more efficiently with newcomers~\cite{Halfaker:2014ky}. Wikidata as Wikipedia's sister project profited from existing experiences in these automation efforts, thus, tools for vandalism detection were highly sophisticated from the beginning~\cite{Sarabadani:2017ia}. %The community relies on this automation and its temporal non-functioning has major consequences for the community~\cite{Geiger:2013fk}. 
However, depending on how this automation is being used, the outcome goes in both directions. The unreflected use of automation can suppress participation of good-faith newcomers~\cite{Halfaker:2012uq}, and on the other hand, recommender systems on Wikipedia can significantly improve editor engagement and content creation~\cite{wulczyn2016growing}. Existing research shows, how the openness of peer production systems, such as the various Wikimedia projects (Wikipedia, Wikidata, etc.) enable researchers to investigate the manifold facets of automation in a real-world setting, and simultaneously support these projects in their goals of providing free high quality content. 

\paragraph{\textbf{Approaches to Algorithm Awareness}}
With regards to related discourses such as on Fairness, Accountability and Transparency (FAT)~\cite{kohli_translation_2018} or Explainable Artificial Intelligence (XAI)~\cite{gunning_explainable_nodate}; algorithm awareness is more aligned to the study of lay's persons experiences of algorithmic systems. As with FAT and XAI, the concerns of the discourse are illustrative of pressing socio-cultural, economic and political needs. However, and similarly to FAT and XAI, algorithm awareness so far suffers from the lack of a methodological definition. Both in terms of design and engineering, implementation of algorithm-aware designs is challenged by two fundamental issues which can be derived from Hamilton and colleagues definition~\cite{hamilton_path_2014}: (1) the perceivability of an algorithm (e.g., results, logic, data) and (2) an actionable mode of representation that leads to informed usage. 
%Open Discussion: Ich finde seltsam, dass Du das Thema der mentalen Modelle vom User komplett rauslässt, aber diese Perspektive ist ja gerade so interessant bei Hamilton, er fragt nämlich danach, wie genau man denn eigentlich bei der Darstellung sein muss und das es nicht gerade um die Beziehung des Menschen zur Technologie geht.

So far, the context of conducted algorithm awareness studies differs greatly. Studies have included, for example, both attempts at reverse-engineering web-based systems such as the Facebook's newsfeed~\cite{eslami_feedvis:_2015} as well as manipulating online peer grading systems~\cite{kizilcec_how_2016}. In the former, Eslami specifies that an algorithm-aware design should provide an actionable degree of transparency to algorithmic processes in order to promote a more informed and adaptive use of a specific system by its users~\cite{eslami_understanding_2017}. In her study, Eslami operationalizes the approach of \textit{seamful design}\footnote{The term seamful design is created as opposite concept to seamless design, where the algorithmic system fades into the background of human's perception.} to display results of the Facebook newsfeed algorithm that usually do not get displayed. 

In the latter, Kizilcec proposes another dimension to algorithm awareness ~\cite{kizilcec_how_2016}: the question of how much transparency of an algorithmic system is actually desirable to ensure understandability and usage. For his study, Kizilcec exposes participants in a peer-graded massive open online course to three kinds of transparency when confronted with their course grades. For each kind of transparency, he asked participants to rate their comprehension of the user interface and to what extent they evaluate the provided information as fair, accurate and trustworthy. These measures provide a first set of measures to empirically study how humans understand algorithmic systems. His results suggest a medium degree of transparency (in this case, textually disclosing the result and logic) as most effective. A high transparency (of the result, the underlying logic and raw peer grading data), he finds, is in fact detrimental to trust in the algorithmic system -- whether or not the received grade was lower than expected. %However, he could show a high degree of correlation (Cronbach's $\alpha$ = 0.83) between the self-report measures.  

A particular focus in algorithm awareness, as well as in XAI and FAT, are the concrete means by which humans may become more informed about algorithmic systems. A frequently deployed solution is the use of textual explanation of algorithmic processes or outputs across all discourses, featuring in contexts such as social media~\cite{rader_explanations_2018}, aviation assistants~\cite{lyons_engineering_2016}, online advertising~\cite{eslami_communicating_2018}, classification algorithms in machine learning~\cite{ribeiro_why_2016} and in online peer grading as discussed above~\cite{kizilcec_how_2016}. The prevalence of this solution may be interpreted as a clear indication for textual explanation being most suitable for establishing algorithm awareness. Within the aforementioned studies, various versions of textual explanations were studied comparatively. For example, even though Kizilcec questioned how much information a user may require, his various conditions of transparency all feature textual, explanatory solutions only~\cite{kizilcec_how_2016}. This may be considered a gap in the discourse. Returning to Hamilton and colleagues, the complexities of contemporary algorithmic systems do not only pose the question of how much humans may need to understand, but also in what way~\cite{hamilton_path_2014}. This suggests, for example, that experimenters should also explore differences between textual explanation of algorithmic logic and interactive, non-declarative solutions in the same context.
%This can be especially seen in the recent emergence of \textit{explainability} and \textit{interpretability} \cite{kohli_translation_2018} as foci in the FAT community.

\paragraph{\textbf{Urgency for Algorithm Awareness}}
Due to the increase of automation on the Web, finding means for a better understanding of algorithms both by experts and lay users is particularly urgent. With algorithms, existing biases may become amplified substantially. In the discourse on recommender systems, bias has been observed as a challenge early on, and a major line of recommender systems research investigates how to avoid popularity bias, i.e., providing recommendations that are already known to satisfy a large number of users~\cite{fleder2007recommender,fleder2009blockbuster}. More recently, several works investigate the explainability of recommender systems~\cite{zhang2014explicit,he2015trirank}.
Even open peer production systems such as Wikidata need to be seen in this context. That is, if there is a pre-existing bias in a knowledge base such as Wikidata, a recommender system may cause this bias to become self-perpetuating. Additionally, encoded bias may spread into the outputs of Wikidata APIs--thereby opaquely influencing the standard in domains that rely on Wikidata services.
In his overview of bias on the Web, Baeza-Yates concludes that an awareness of bias (whether algorithmic or cultural) is the primary precondition for designers and engineers to mitigate potentially negative effects on users~\cite{baeza-yates_bias_2018}. The developer perspective as advanced by Baeza-Yates suggests that an engineering solution may be found with the potential to eliminate bias, whether by way of analyzing biased tendencies in the data used by a Web platform or running of extensive A/B-testing of subgroups \cite{kohavi_controlled_2009}.

However, as repeatedly noted by Wiltse and Redstr\"om, the complexity of algorithmic systems in the modern Web troubles this suggestion. In their words, the Web is populated not by clear developer-client relations, but by \textit{fluid assemblages}, i.e. socio-technical configurations that change in various context of use \cite{redstrom_press_2015,wiltse_wicked_2015}. Bias, therefore, is not necessarily a definitive phenomenon for either human or machine. Accordingly, counting on purely technical solutions to eliminate bias needs to be up for debate. Instead, and as called for by various researchers from algorithm awareness, FAT and XAI, empirical studies that provide insights into how algorithmic systems (and the biases encoded therein) may be made more transparent.

In the next section, we introduce the context of the open peer production system Wikidata, in which our use case, Recoin -- a property recommender-system -- is used. 

%******************************************************************************************************************************************************
\section{Recoin: Property Recommender} 

Wikidata is an open peer production system~\cite{Vrandecic:2014hl}. Its structured data is organized in entities, where two types of entities exist: items and properties. Items represent real-world objects (individuals) or abstract concepts (classes). 

Each item is described by statements; for example, the entity \texttt{Q1076962} represents the human \textit{Chris Hadfield}. Each item is described by statements that follow a \textit{subject-predicate-object} structure (e.g., \textit{Chris Hadfield} (\texttt{Q1076962}) ''is instance of'' (\texttt{P31}) ''human'' (\texttt{Q5})). Thus, a property, i.e. predicate, describes the data value, i.e. object, of a statement. In October 2018, the community has more than 200k registered contributors with 19K active on a monthly base. They have created more than 570m statements on more than 50m entities. 

Even though Wikidata was founded to serve as a structured data hub across all Wikimedia projects, today, it is utilized for many other purposes; for example, researchers apply Wikidata as authoritative for interlinking external datasets, such as for gene data~\cite{burgstaller2016wikidata} or digital preservation data~\cite{thornton2017modeling}, or companies use Wikidata's knowledge graph for improving their search results, such as Google or Apple. A significant issue for Wikidata's community is consequently the quality of the data. Data quality is a classical problem in data management, however, in peer production settings such as in Wikidata, data quality assessment is complicated because of the continuous, incremental data insertions by its users, the distributed expertise and interest of the community, and the absence of a defined boundary in terms of its scope. Over the past years, the community has introduced many tools that address this challenge, that range from visualizing constraint violations to de-duplication tools and translation tools. One of these tools is Recoin that we present in more detail in the next section. 

%********************************************************
\subsection{Technical Design}

Recoin is a recommender system for understanding and improving the completeness of entities in Wikidata~\cite{ahmeti2017assessing,balaraman2018recoin}. A main motivation for implementing Recoin is Wikidata's openess, since it allows anyone to add nearly any kind of entities - items and properties. %The community has just started to employ  consistency constraints programmatically. Because of its openess, 
The latter led to a huge space of possible properties (4,859 properties as of July 2nd, 2018), with many applying only to a very specific context (e.g., ''possessed by spirit'' (\texttt{P4292}) or ''doctoral advisor'' (\texttt{P184})). Consequently, even experienced editors in Wikidata may lose track of which properties are relevant and important to a given item which might hinder them to improve data quality in Wikidata~\cite{razniewski2017doctoral}. %For specific contexts, editors would in cases manually define properties relevant for a class.\footnote{\label{note1}Examples: https://www.wikidata.org/wiki/Wikidata:WikiProject\_Q5/lists/ \linebreak riders\_and\_their\_horse and https://www.wikidata.org/wiki/Wikidata: \linebreak Wikivoyage/Lists/Embassies}\clmb{Ich verstehe die Relevanz des letzten Satzes nicht. Braucht es diesen wirklich? Wenn ja, warum?}

\begin{figure}[t]
\begin{center}
\includegraphics[width=6cm]{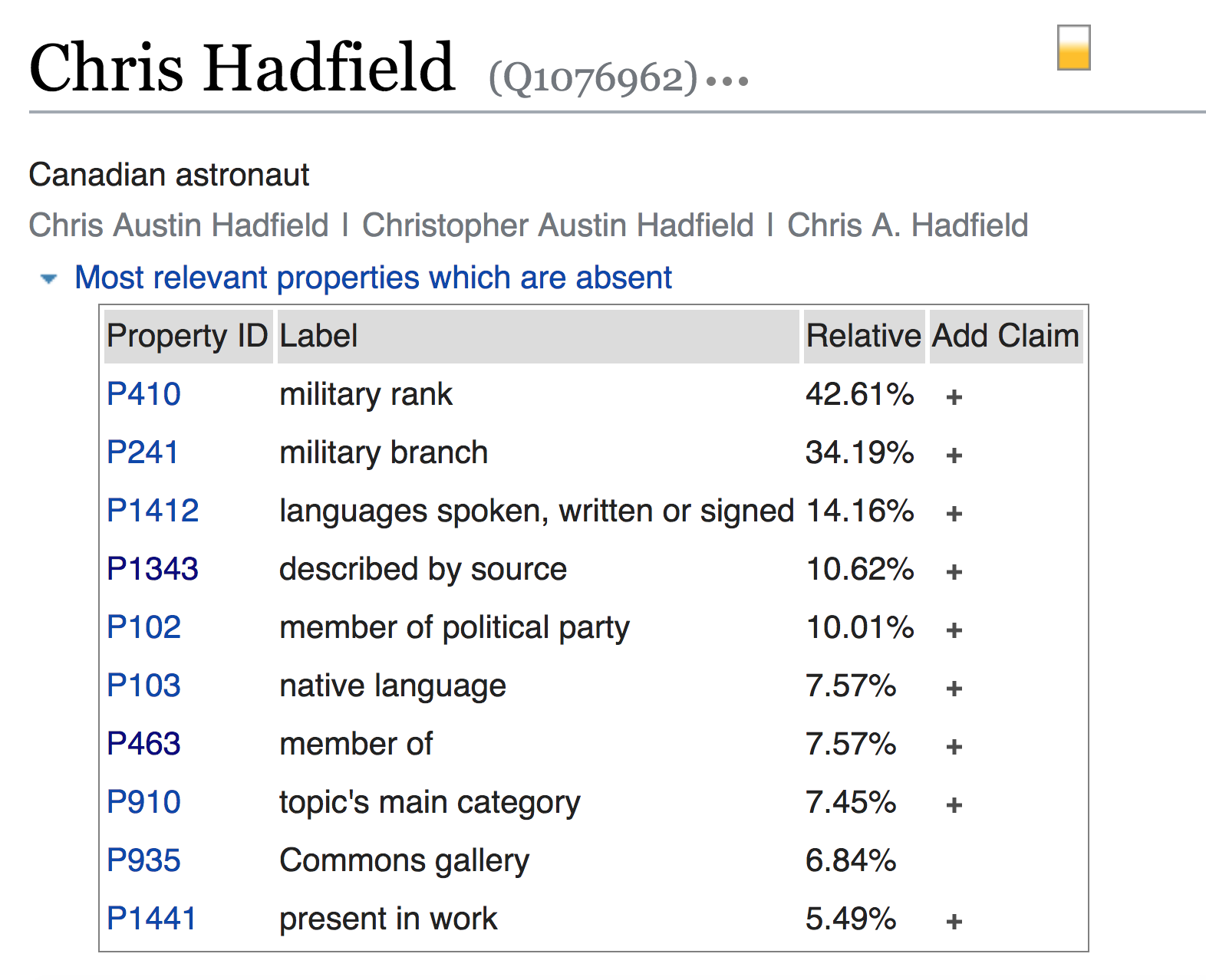}
\caption{Recoin for the astronaut Chris Hadfield.}
\label{fig:recoin}
\end{center}
\end{figure}

Recoin is a gadget - an interface element - on Wikidata\footnote{Further information is available at \url{https://www.wikidata.org/wiki/Wikidata:Recoin}.}. A visual indicator informs a person about the relative completeness of an item and, moreover, it provides an editor with concrete recommendations about potentially missing properties on this item.
%Recoin is an interface element - a gadget - that aims to (i) recommend relevant properties for an item to editors, and (ii) to visualize the relative completeness of an item. 
Figure~\ref{fig:recoin}) shows the gadget on an item page on Wikidata. A visual indicator (icon on the top right) shows a color-coded progress bar that has five levels ranging from empty to complete. On the top of an item page, the recommendations are provided in an expandable list that shows up to ten of the most relevant missing properties. 

The idea of relative completeness is motivated by the situation that in absolute terms, measuring the completeness of an item is impossible in an open setting. The relative completeness, thus, considers completeness in relation to other, similar items. The relatedness function of Recoin considers two items as similar if they share the same class\footnote{An exception are items that are instance a of the class \emph{human}. In this case, the class ''occupation'' is used.}. The visual indicator of Recoin should not be understood as an absolute statement, i.e. level 5 (=complete) means, that all possible statements a given on the item page, but should rather be interpreted as a comparative measure, i.e. the statements on this item are more complete than on similar items. 

The completeness levels in Recoin are based on thresholds that are manually determined. It considers the average frequency of the 5 most frequent properties among the related ones to consider an item as \emph{most complete} (0\%-5\% average frequency), \emph{quite complete} (5\%-10\% average frequency), and so on. Furthermore, each user is shown 10 recommendations in order to avoid an overwhelming user experience. 

%Recoin's architecture is organized in four components (cp. Figure~\ref{fig:recoin-flow}). When a person request a Wikidata item page, the frontend (green) calls the backend (yellow), which in turn uses a database (blue) plus live data from Wikidata from the SPARQL endpoint (red).  in order to determine the most relevant missing properties from the given entity. It then uses their frequency to compute the completeness level, as described before.

%\begin{figure}[t]
%\begin{center}
%\includegraphics[width=8cm]{recoin-flow}
%\caption{Information flow inside Recoin.}
%\label{fig:recoin-flow}
%\end{center}
%\end{figure}

%********************************************************
\subsection{Need for Algorithm Awareness}
\label{sec:needalawarness}

As of September 25, 2018, Recoin is enabled by 220 editors on Wikidata\footnote{Further information are provided on the following page \url{https://www.wikidata.org/wiki/Special:GadgetUsage}.}, who created, based on Recoin's recommendations, 7,062 statements on 4,750 items.
% TODO: Diese Zahlen müssen noch einmal überprüft werden.
Even though Recoin is a straightforward approach for improving data quality on Wikidata, editors are hesitating to apply Recoin. Moreover, after persons used Recoin, they have raised a number of concerns. Based on existing discussions on Recoin's community page\footnote{\url{https://www.wikidata.org/wiki/Wikidata_talk:Recoin}} and on the mailing list, we identified three typical issues. 

Editors, for example, posed questions regarding the \textit{scope of the recommender}: \emph{``Not sure if Recoin looks at qualifiers \& their absence; if not, might this be something to think about?''}. The information provided by Recoin hindered editors to understand which data was being used to compute the recommendations. In another case, an editor was wondering about Recoin's underlying algorithm: \emph{``Something weird going on on Jan Jonker (Q47500903), stays on least complete.''}. In this case, the unchanging visual indicator of Recoin caused the user to \textit{question the functionality} of Recoin. Another user was concerned about the provided recommendation and its suitability for specific items: \emph{``How is Property:P1853 "Blood type" on this list, is that relevant (or even desirable) information for most people?''}. The user was not able to include its personal preferences - world view - in Recoin's recommendation. 

However, the third typical issue exemplified by Wikidata's editors raises a more genuine concern over the \textit{impact of Recoin on an already biased knowledge base} (e.g., the predominance of the English language~\cite{kaffee_glimpse_2017}). One editor stated: \emph{
``This tool has its attractions but it does entrench cultural dominance even further as it stamps "quality" on items. The items with the most statements are the ones that are most likely to get good grades. Items from Indonesia let alone countries like Cameroon or the Gambia are easily substandard.''}.\footnote{The corresponding thread can be found on Wikidata's mailing list archive~\url{https://lists.wikimedia.org/pipermail/wikidata/2017-December/011576.html}.} 

On a surface read, this quote further substantiates the misunderstood nature of Recoin's function, as it is not intended as a unilateral absolute grading of the completeness of a particular item, but rather as a comparative tool that recommendations depend on the activities of the editors on similar items. However, and much more significant, the concern raised about cultural dominance is a very contemporary problem in algorithmic system design. Recoin fails to address this concerns by its current design and mediated function. In other words, the cultural bias in the recommended properties, even if not intended, seem to affect the usage of Recoin.

%For example, while there are 178 active Wikidata users of Recoin, the top five users account for more than half of the 7061 edits performed with Recoin. This type of activity bias skews the relevance of the properties, potentially leading to an algorithmic bias in Recoin due to the self-perpetuating recommendation of adding the relevant properties \cite{baeza-yates_bias_2018}.\clmb{Jesse: Was genau ist hier der Punkt? Kann das noch weiter ausgeführt werden?}

Based on these insights, we wanted to better understand how a re-design of Recoin that considers algorithm awareness by focusing on explainability and interactivity can address the aforementioned issues. As opposed to existing research in this context, for example carried out by Eslami et al.~\cite{eslami_feedvis:_2015}, we do not require methods such as reverse engineering to understand the algorithmic system we are dealing with on a technical level. This knowledge is key to understanding the intricacies of the Web platforms today; as the ways in which an algorithm operates within a larger socio-technical context arguably also shapes the extent to which humans can or should be aware of it. Therefore, with an openly available recommender system in an open peer production system, we can conduct experiments that are closely tied to the actual practice of Wikidata editing activities, i.e., we can reflect on the technical and the social system similarly.

In the following section, we introduce our experimental setup that help us to examine the impact of varying degrees of explainability and interactivity of the UI of the recommender system on humans. Following the concept of Recoin, our experiment featured a task of data completion. By measuring the interactions of participants with various designs during a data completion task and by eliciting self-reports, we sought to understand which design measures increased task efficiency while at the same time were most effective in increasing understanding of the algorithmic system.

\begin{figure}[t]
\begin{center}
\includegraphics[width=6cm]{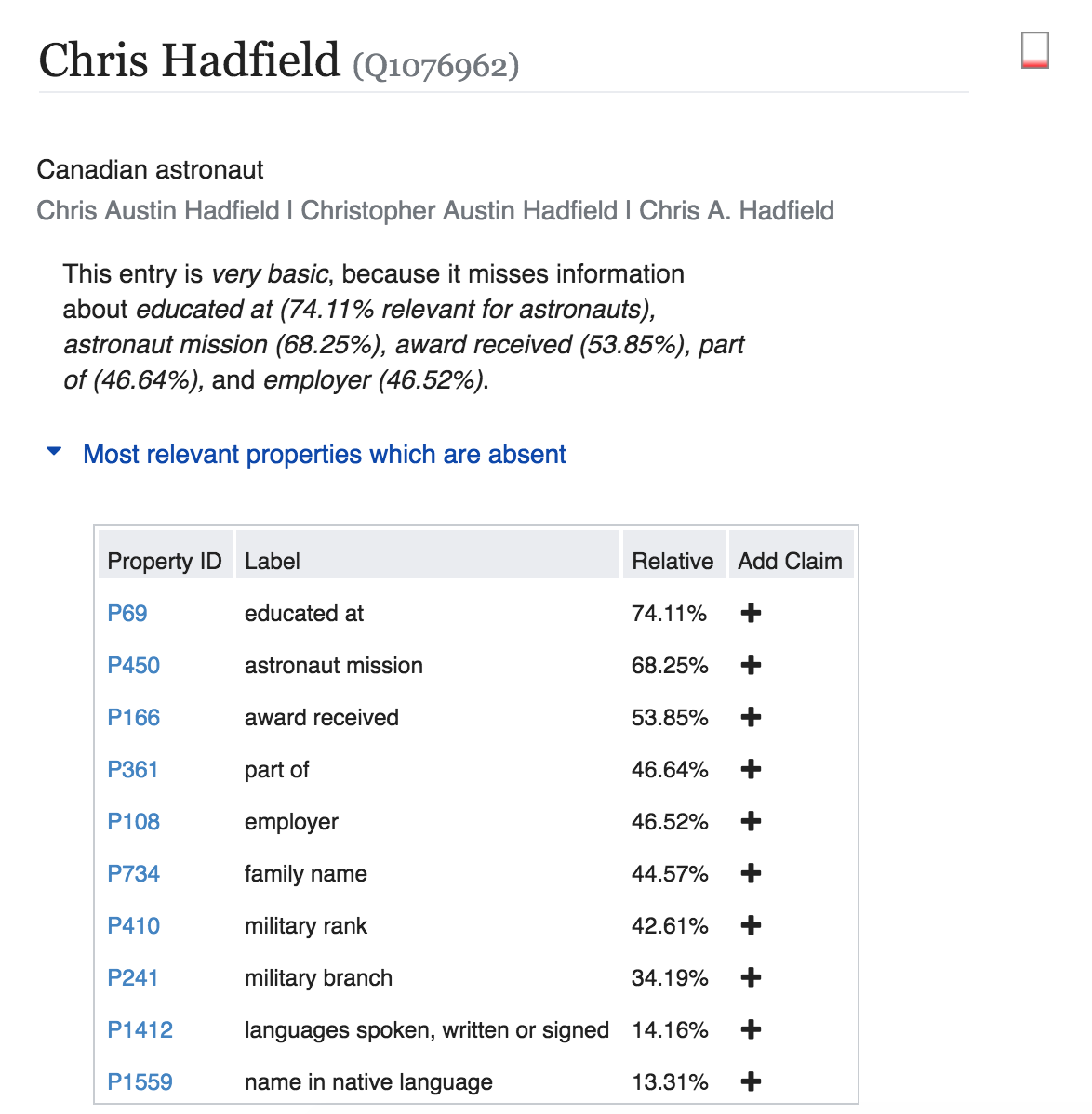}
\caption{Recoin with Explanation \textit{(RX)}.}
\label{fig:recoin-x}
\end{center}
\end{figure}

%**************************************************************************************************************************************************************************
\section{Experimental Setup}

Informed by previous research, we designed two alternative UIs where each represents another degree of explainability and interactivity. Each user interface extends or replaces the previous version by specific design elements. In the following, we differentiate the original Recoin design (R1), the textual explanation design (RX), and the interactive redesign (RIX). The RX design is mainly inspired by previous work, from which we adapted the explanation design~\cite{kizilcec_how_2016}. The RIX design follows an interactive approach, where the user can interact with the outcome of the algorithm, thus, can explore how the algorithm outcome changes based on specific settings. The three UI designs are used in five experimental conditions (C1 to C4), supplemented by a baseline where participants only used the regular Wikidata interface. We explain these conditions in more detail in one of the following paragraphs.

Based on these designs, participants should solve the same task across all conditions: adding data to a Wikidata item. We recruited 105 participants via Amazon Mechanical Turk (MTurk)\footnote{For more information please check \url{https://www.mturk.com/}.}, whereby each participant had a minimum task approval rate of 95\%, and a minimum amount of HITs of 1,000. Each participant received USD 3.50 (equivalent to USD 14.00 hourly wage) for full participation. We recruited only U.S. participants to reduce cultural confounds. We randomly and evenly distributed participants over our five conditions (i.e. 21 participants), also ensuring that no participant could re-do the task or join another condition by associating participants with qualifications. Each participants was given 10 minutes for task completion. 

In each condition, participants went through the same general procedure during task completion. At first, we provided a brief on-boarding, then we provided a task briefing. After the study participant carried out the task, she had to fill out an explicative self-report which contained the dimensions comprehension, fairness, accuracy and trust. Additionally, all participants obtained a task completion score, which we correlated with server activity to ensure that our final study corpus featured no invalid contributions. All data and results of our study will be available under an open license\footnote{Omitted for review.}. 

In the following, we outline the design decisions that have led to our three designs for Recoin in more detail. Then, we describe the task and the experimental design. 

%**************************************************************************************************
\subsection{Design Rationales}

In the following, we describe each UI approach of the recommender Recoin in more detail. For each user interface, we provide a corresponding visual representation.

\subsubsection{Recoin User Interface R1}
The original design of Recoin (cp. Figure~\ref{fig:recoin}) was primarily informed by existing UI design practices in Wikipedia. The status indicator icon was chosen to mirror the article quality levels on Wikipedia, such as "Good article" or "Featured article" \footnote{For more information we refer to \url{https://en.wikipedia.org/wiki/Wikipedia:Good_articles} and \url{https://en.wikipedia.org/wiki/Wikipedia:Featured_articles}.}. The used progress bar was motivated by existing visualizations in Wikipedia projects.\footnote{Examples are \url{https://www.wikidata.org/wiki/Wikidata:WikiProject\_Q5/lists/riders\_and\_their\_horse} and \url{https://www.wikidata.org/wiki/Wikidata:Wikivoyage/Lists/Embassies}.} 

Some parameters to represent the results of the Recoin recommender were determined without further considerations, such as the thresholds that represent the five levels of completeness. 

\begin{figure}[t]
\begin{center}
\includegraphics[width=8.2cm]{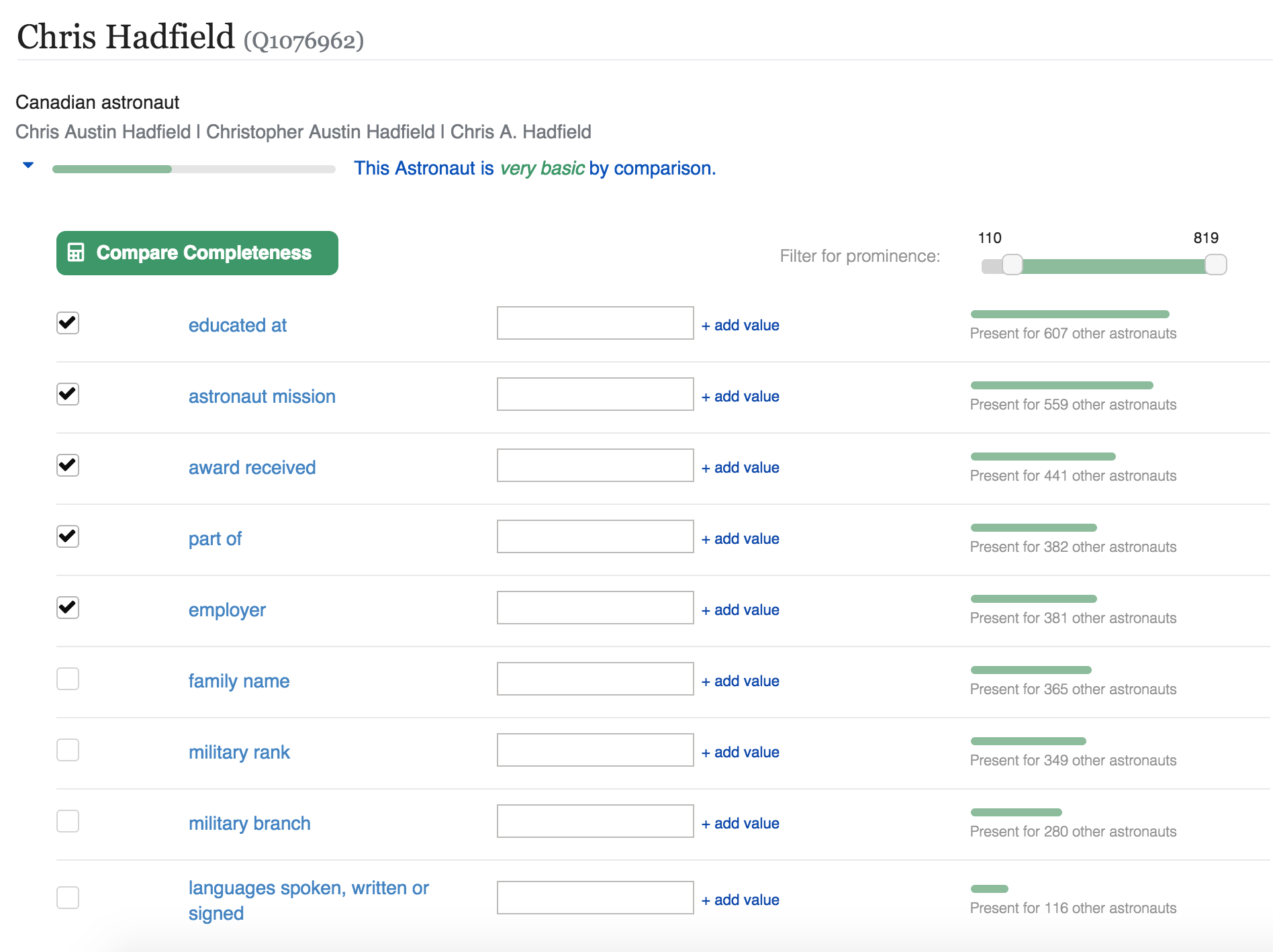}
\caption{Recoin Interactive Redesign \textit{(RIX)}.}
\label{fig:recoin-ix}
\end{center}
\end{figure}

\subsubsection{Recoin User Interface RX}
Textual explanation of algorithmic logic is a wide-spread measure in the related work, and has been deployed in contexts such as social media~\cite{rader_explanations_2018}, aviation assistants~\cite{lyons_engineering_2016}, on-line advertising~\cite{eslami_communicating_2018} and classification algorithms in machine learning~\cite{ribeiro_why_2016}. For our design of RX, we drew inspiration from Kizilcec, who tested three states of transparency to understand algorithm awareness in on-line peer grading~\cite{kizilcec_how_2016}. Since the algorithm's function can be compared to Recoin (i.e., a rating algorithm), we adapted the format of Kizilcec's best solution and added a textual explanation to Recoin's user interface that describes the logic behind Recoin's calculation (cp. Figure~\ref{fig:recoin-x}). 

\subsubsection{Recoin User Interface RIX}
Our interactive user interface (RIX, cp. Figure~\ref{fig:recoin-ix}) is based on insights gained from user feedback from Recoin's current users (cp. Section~\ref{sec:needalawarness}) and from the philosophy of technology as discussed by Verbeek~\cite{verbeek_what_2006}. Concerning the latter, we posit that Recoin \textit{actively} transforms the relationship an editor has with Wikidata and the entities therein. Through Recoin, Wikidata items that formerly were objects containing knowledge are now also objects that are rated. Technically, this rating is not an indication of absolute qualities, but one of community-driven standards, i.e. how the Wikidata community currently views a specific class of items.

However, as illustrated in the various responses to Recoin, this mediation is not adequately communicated by Recoin's current design. Furthermore, in reflecting on Recoin with the original developers, we found that the comparative parameter of dividing the relevancy of the top five properties was arbitrarily chosen. In line with Mager~\cite{mager_internet_2018}, we consider transparency for this result of developer decision-making as essential. 

We operationalized these insights for RIX by considering how the community-driven aspect of Recoin could not only be displayed, but made interactively explorable. 
To this end we (1) included a reference to the class of the displayed entity (e.g., ''astronaut'' in our running example) in the drop-down title. This was designed to convey that this particular item is rated based on its class. Next, we augmented the drop-down itself extensively. 
We (2) substituted the relevance percentage with a numerical explanation for each suggested property (e.g., a relevance for the property 'time in space' of 67.03\% means that 549 out of 819 astronauts have this property). In contrast to a percentage, it was our intuition that relating to the class would highlight the community-driven aspect of Recoin. 
To strengthen this aspect further, we (3) included a range slider which allows filtering properties based on their prominence in the class (i.e., compare this entity based on their occurrence in a minimum/maximum of \textit{n} astronauts). 
Finally, we offered a way for directly interacting with Recoin's calculation: we (4) allowed our participants to reconfigure the relevancy comparison by (de-)selecting individual properties. Thereby, we wished to show that relevancy can be a dynamic, community-driven attribute in this algorithmic system.

\begin{figure}[t]
\begin{center}
\includegraphics[width=8cm]{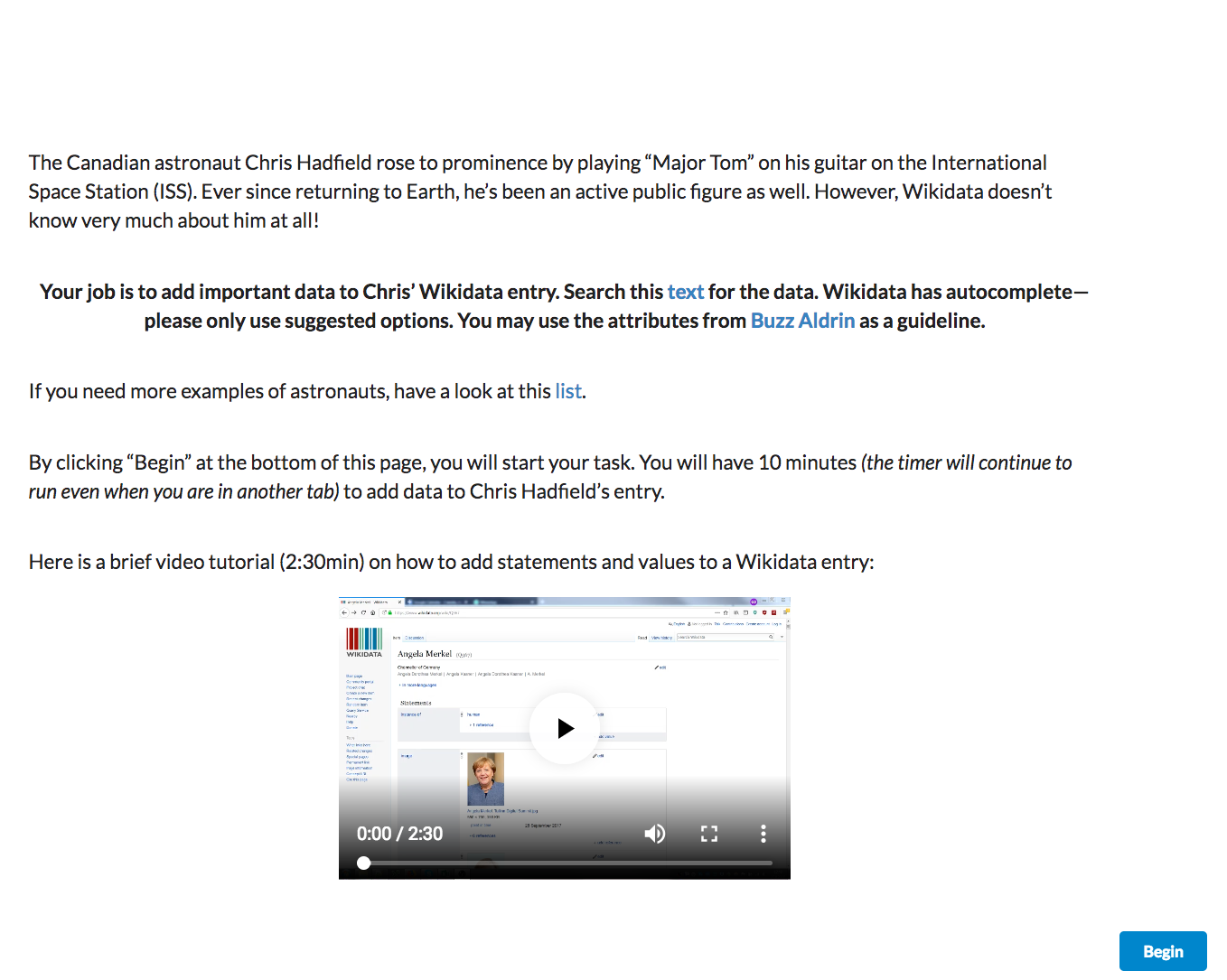}
\caption{Briefing page, with material added and resources for manually carrying out Recoin's functions and tutorial.}
\label{fig:briefing}
\end{center}
\end{figure}

%******************************************************************************************
\subsection{Task}

For the study participants, we defined a typical editing task on Wikidata. We presented each study participant with a copy of Wikidata's user interface, to provide a most realistic task setting. First, the participants received a brief on-boarding for Wikidata, and, depending on the condition, for Recoin as well. Participants then proceeded to the task briefing page (cp. Figure~\ref{fig:briefing}). The participants were asked to add further properties and data to a Wikidata item. Additionally, we supplied participants with a short video tutorial that explained how properties can be added to an item on Wikidata. 

In each condition, the Wikidata item to be edited was \textit{Chris Hadfield}, a Canadian astronaut\footnote{The original page is provided here: \url{https://www.wikidata.org/wiki/Q1076962}.}. This item was chosen because it has a number of missing statements that are easily retrievable, and on the other hand, the item describes an astronaut who is probably well-known by our study participants which are U.S. based. Additionally, the occupation of astronauts was thought to be relatively neutral, as opposed to, for example, politicians or soccer players. 

We provided study participants with source material for the task composed of comparatively relevant and irrelevant pieces of information about \textit{Hadfield}. We also supplied a link to a very detailed Wikidata item with the same occupation, the US-American astronaut \textit{Buzz Aldrin}\footnote{\url{https://www.wikidata.org/wiki/Q2252}.}, and a link to a Wikidata query of the occupation ''astronaut''\footnote{Please check \url{http://tinyurl.com/ycnh3q37}.}, both with the intention to allow study participants to compare the given item with other items, i.e. we encouraged our participants to perform the functionality of Recoin manually. 

In addition, we provided a short video tutorial on how to add statements to Wikidata items. Following the task briefing, participants could choose to commence the task, which lead to the reconstructed Wikidata page for \textit{Hadfield}. Within a 10 minute limit, participants could then add statements to the item.

We randomly assigned each participant to one of the conditions. Once the 10 minutes passed, participants were alerted that time was up, and that they should proceed to the self-report. Here, participants were confronted with a grade (from A-F) of their task. This grade was calculated through the difference in completeness before and after participants added information to the Wikidata item (e.g., when a participant additions increased the relative completeness of \textit{Hadfield} by more than 20\% but less than 30\%, they received a "B").

In correspondence to this grade, participant's were asked to rate their comprehension (5-point Likert scale), feelings of accuracy, fairness and trust (7-point Likert scale) of the recommender system. Again, due to substantial methodological and contextual similarities to Kizilcec's online study~\cite{kizilcec_how_2016}, we adopted the aforementioned measures to our study. Participants were also asked to expand on their ratings using free text fields. Upon submitting their ratings, participants were returned to the MTurk platform.

\begin{table}[t]
\begin{tabularx}{\columnwidth}{XL{6cm}}
\toprule
\textbf{Relevance} & 
Difference of the completeness value of the item before and after task completion. \\ \hline
\textbf{Usage} & 
Number of times the recommender Recoin was used during task completion. \\ \hline
\textbf{Compre\-hension} & 
To what extent do you understand how your task has been graded? \textit{(1) No understanding at all to (5) Excellent understanding.} \\ \hline
\textbf{Fairness} & 
How fair or unfair is your grade? \textit{(1) Definitely unfair to (7) Definitely fair.} \\ \hline
\textbf{Accuracy} & 
How inaccurate or accurate is the grade? \textit{(1) Very inaccurate to (7) Very accurate.} \\ \hline
\textbf{Trust} & 
How much do you trust or distrust Wikidata to fairly grade your task? \textit{(1) Definitely distrust to (7) Definitely trust.}\\
\bottomrule
\end{tabularx}
\caption{Overview of measures employed in our online experiment.}
\label{measure-table}
\end{table}

%******************************************************************************************
\subsection{Study Design}

We conducted a between-subject study with five conditions. In the following, we define each condition and explain each measure we collected during the study. 

\subsubsection{Conditions}

The first three conditions (baseline, C1, C2) were designed to test usage and understanding of the current version of Recoin, i.e. R1. We then proceeded to test the collected baseline against textual explanation (C3 with RX) as found in related work \cite{kizilcec_how_2016,gunning_explainable_nodate,phillips_interpretable_2018} and a redesign motivated by the shortcomings found therein (C4 with RIX). By comparing the results of the conditions, we aimed to gather insights on how design impacted human understanding of Recoin's function. 

All conditions are described in more detail next, followed by a description of the collected measures.
\begin{itemize}
\item \textit{Baseline}: Participants can add data on a Wikidata item \textit{without} Recoin being present in the user interface.
\item \textit{Condition 1}: Participants can add data on a Wikidata item \textit{with} Recoin (R1) being present in the user interface.
\item \textit{Condition 2}: C1 but Recoin is mentioned during the on-boarding process.
\item \textit{Condition 3}: C2 but Recoin (RX) with explanation interface. 
\item \textit{Condition 4}: C2 but Recoin (RIX) with interactive interface.
\end{itemize}

\subsubsection{Task Measures} 
\textbf{Relevance}: As the improvement of data quality is the primary goal of Recoin, we wanted to ensure that we understood how each condition affected the change in completeness; independent of the quantity of contributions. Thus, we defined the metric \textit{relevance} as our dependent variable. Relevance is defined as the difference of the completeness values of Recoin before and after a participant added properties to the item.

\textbf{Recoin Usage:} As a recommender system, it is particularly important to understand how each condition (aside from the baseline) affected the number of times Recoin was used directly to add information to an item. This is expressed by the measure \textit{usage} which serves as a dependent variable.

\textbf{Time:} We fixed the time participants can add properties to an item to ensure that our conditions are comparable. The measure \textit{time} serves as our control variable. 

\textbf{Demographics:} All study participants were recruited via MTurk. While we assume that the majority of participants are US-Americans, we did not further specify our demographics. Thus, as is typical, demographics were our covariates.  

\begin{figure}[t]
\begin{center}
\includegraphics[width=7cm]{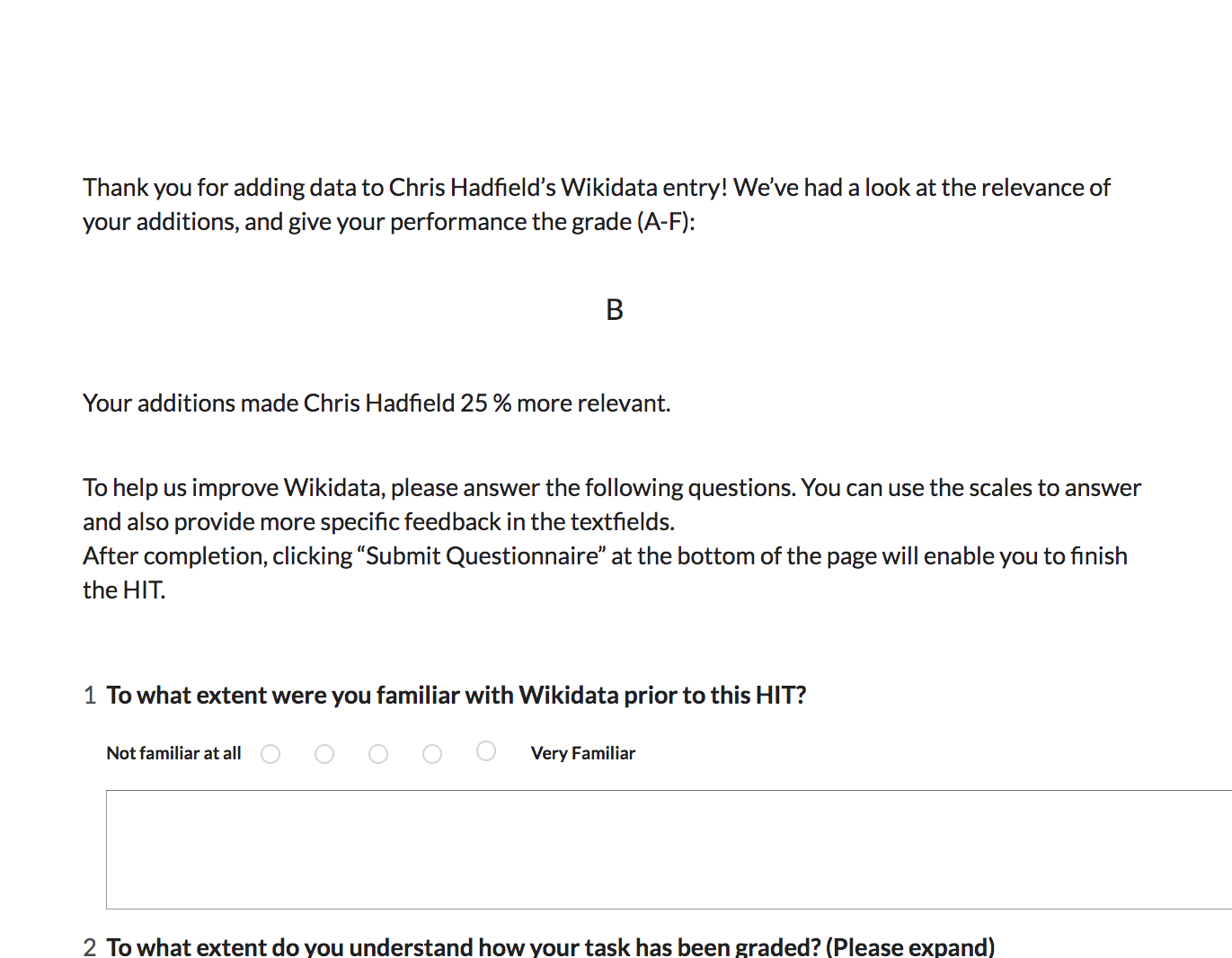}
\caption{Questionnaire after adding properties that lead to an increase in relevance of 25\%.}
\label{fig:questionnaire}
\end{center}
\end{figure}

\subsubsection{Self-Report Measures} Upon completion of the task, participants were directed to a self-report page (cp. Figure~\ref{fig:questionnaire}). The page prominently featured a grading of the participant task performance, which was calculated by normalizing the average comparative relevance, i.e. Recoin's assessment, of each contribution per participant. We graded the participants performance in their task (A-F) in order to elicit a reaction to the task even if participants did not notice, use or understood Recoin. We purposefully designed this grading to encourage study participants to reflect on the task, for example a participant may receive the grade F despite many additions to the item. 

Furthermore, we included ratings with the four factors from previous research on algorithm awareness~\cite{kizilcec_how_2016}: comprehension (5-point Likert scale), accuracy, fairness and trustworthiness (7-point Likert scale) of the algorithmic system. These measures should be strongly correlated according to the procedural justice theory in the related work. Low ratings of all measures, for example, would stem from violated expectations in an outcome~\cite{kizilcec_how_2016}. %We wanted to see how participant performance was related to Recoin's assessment, and whether an grades would allow us to correlate the quantitative results of our algorithm aware designs and self-reported perceptions more strongly. 
%Furthermore, we wished to test the relationship of the metrics used by Kizilcec in our use case, thereby hoping to gain an understanding of their usefulness for experiments on algorithm awareness. 
We asked all participants in addition, to expand on their ratings via text-fields for collecting qualitative data as well.

%\subsubsection{Miscellaneous} We collected participant interaction data (e.g. mouse movements, time needed, number of clicks) to broaden our understanding of how specific design choices of the UI approaches affected the results of each condition.\footnote{We employed the Javascript Library hotjar.js (\url{https://www.hotjar.com/}).} 

\begin{table}[t]
\begin{tabularx}{\columnwidth}{XC{1.5cm}C{2cm}C{2cm}}
\toprule
\textbf{Condition} & \textbf{\# All Edits} & \textbf{\# Recoin Usage} &  \textbf{SD Recoin Usage}\\ 
\midrule
\textbf{Baseline} & 249 & - & - \\
\textbf{C1} & 319 & 61 & 3.10 \\
\textbf{C2} & 382 & 91 & 8.56 \\
\textbf{C3} & 301 & 55 & 3.50 \\
\textbf{C4} & 281 & 71 & 4.25 \\ 
\bottomrule
\end{tabularx}
\caption{Number of edits, i.e. contributions in each condition, with the number of Recoin usage and standard deviation.}
\label{edittable} 
\end{table}

%**************************************************************************************************
\subsection{Hypotheses}
%\clmb{Jesse: Ich finde die Hypothesen sehr schwierig, denn sie kommen ein wenig aus dem Nichts. Erst einmal braucht es noch ein wenig mehr Erklärung, jeweils vor jeder Hypothese, ansonsten finde ich sie wirklich unverständlich. Mir ist aus den Worten nicht recht klar, worauf Du Dich immer beziehst. Kannst Du das bitte ergänzen.}
%\clmb{Jesse: Du musst die Formulierung Deiner Hypothesen ändern: A hypothesis both defines the variables involved and the rela- tionship between them, and can take many forms: A causes B; A is larger, faster, or more enjoyable than B; etc. - siehe dazu Ways of Knowing in HCI. (n.d.). Ways of Knowing in HCI. Seite 196!!!}
For our hypotheses, we were interested in testing the impact of Recoin on data completeness. Having provided participants with equal opportunities to add relevant data to the item \textit{Hadfield}, we examined whether Recoin improves the completeness of an item or not. 

Based on our analysis of the status quo (cp. Section~\ref{sec:needalawarness}), we did not expect study participants to actively use Recoin which lead to the following hypothesis: \textbf{$H_{1}$:} Using Recoin does not lead to significantly higher relevance in terms of data completeness.

Based on the discussed literature on algorithm awareness, we assume that an user interface that conveys explainability and interactivity of the underlying recommender system leads to higher usage rates: \textbf{$H_{2}$:} The interface design of Recoin impacts the number of time participants used Recoin.

Furthermore, we assumed that the effectiveness of algorithm aware designs would be captured most succinctly by the comprehension measure, which would accordingly allow us to distinguish the impact of the RX and RIX designs. Given the results  of textual explanation employed in related work (cp. Section~\ref{sec:relatedwork}), we therefore hypothesized that: \textbf{$H_{3}$:} A textual explanation of the algorithmic logic leads to higher comprehension than the interactive redesign.

Finally, to gain insights on methodological procedure, we  sought to test the experimental self-report measures employed by Kizilcec~\cite{kizilcec_how_2016}. According to this research, the self-report measures should exhibit a high degree of correlation (Cronbach's $\alpha$ = 0.83). We therefore hypothesized: \textbf{$H_{4}$:} The correlation of self-report measures for textual explanation solutions will equally hold for testing the interactive solution.

%***************************************************************************************************
\subsection{Results}

\begin{table}[b]
\begin{tabularx}{\columnwidth}{XC{0.9cm}C{0.7cm}C{0.95cm}C{0.7cm}C{0.7cm}C{0.8cm}}
\toprule
\textbf{Condition} & \textbf{Grade} & \textbf{Rel.} & \textbf{Comp.} & \textbf{Fair.} & \textbf{Acc.} & \textbf{Trust} \\ 
\midrule
\textbf{Baseline} & C & 11 & 2 & 4 & 4 & 4 \\
\textbf{C1} & C & 15 & \textbf{3} & 4 & 5 & 4 \\
\textbf{C2} & C & 19 & \textbf{3} & 5 & 5 & \textbf{5} \\
\textbf{C3} & \textbf{B} & 20 & \textbf{3} & 4 & 4 & 4 \\
\textbf{C4} & \textbf{B} & \textbf{21} & \textbf{3} & \textbf{6} & \textbf{6} & \textbf{5} \\ 
\bottomrule
\end{tabularx}
\caption{Median values for (1) task performance: \textit{Grades} dependent on the increase of \textit{Relevance}; (2) self-report: \textit{Comprehension} (1 - 5), \textit{Fairness}, \textit{Accuracy} and \textit{Trust} (1-7).}
\label{mediantable} 
\end{table}

We recruited 21 participants for each condition ($n = 105$). Overall, we received 1,532 edits (cp. Table~\ref{edittable}), with participants in the \textit{C2}-condition providing the most. In the \textit{C4}-condition, our interactive redesign, participants used Recoin most frequently, with more than half (61.09\%) of participants adding data via the Recoin interface at least once. This condition also included the most relevant contributions, with a median increase in completeness for the Hadfield item of 21\%. The median values of task performance, i.e., received grade and average increase in completeness, as well as the ordinal Likert-scales from the participant self-report, i.e., comprehension, fairness, accuracy and trust, can be seen in Table~\ref{mediantable}. 

We expected only a small amount of qualitative data. However, we found that displaying a grade in the self-report provided a highly effective trigger. Overall, 82 of our 105 participants chose to expand on their self-reported ratings via the provided text fields. This allowed us to probe participant statements for insights on specific subjective perspectives.

In the following, we show the results of our analysis for each hypothesis by using the Kruskal-Wallis test for ordinal data and ANOVA for numerical data. We report the results of the algorithm awareness measures with Spearman correlation tests. Finally, we provide findings of our qualitative analysis of participant statements.

\begin{figure}[t]
\begin{center}
\includegraphics[width=7cm]{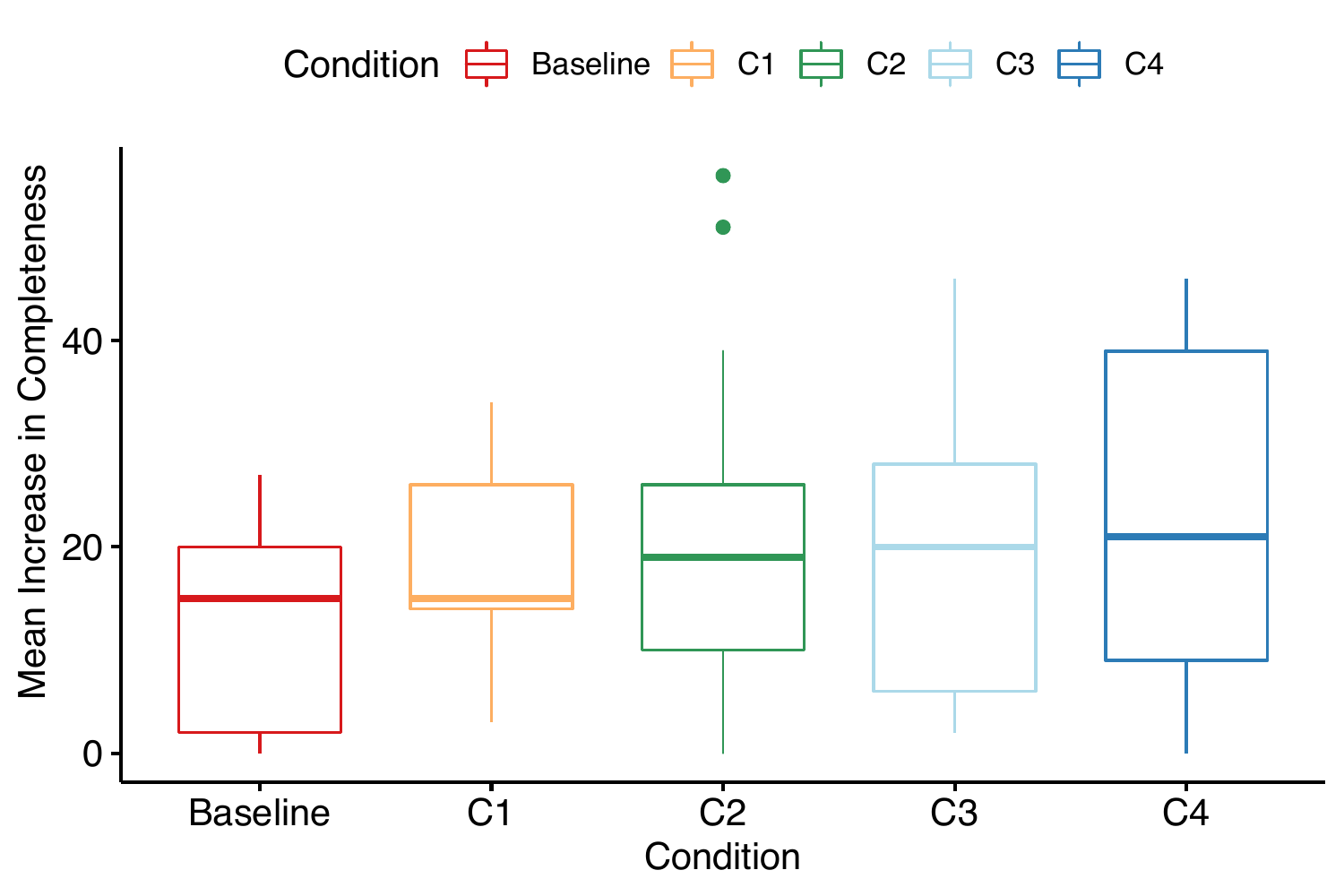}
\caption{Boxplot of the mean increase of completeness per condition.}
\label{fig:rel-con}
\end{center}
\end{figure}

\subsubsection*{$H_{1}$: Using Recoin does not lead to significantly higher relevance in terms of data completeness.} We reject this hypothesis. An increase of comparative relevance for the Hadfield item is highly dependent on using Recoin at least once ($p_{rel,rec} < 0.001$). Additionally, when looking at the increase of comparative relevance per participant as a function of the numbers of additions made via Recoin, we can see that the most significant difference occurs around 7 additions made (\textit{p\textsubscript{rel,numUse}} = 0.02). This shows that Recoin is highly efficient, as adding a majority of the ten recommended properties should lead to the highest increase in relevance.

\subsubsection*{$H_{2}$: The interface design of Recoin impacts Recoin usage.} Even though the redesign (C4) slightly outperformed the other conditions in terms of the goal of the set task, we could not find any significant difference for the number of additions made via Recoin between C1, C2, C3, or C4 ($p = 0.74$). Therefore, we cannot confirm this hypothesis with statistical significance.

\subsubsection*{$H_{3}$:  Textual explanation of the algorithmic logic leads to higher comprehension than the interactive redesign.} We could not find statistically significant differences between ratings of comprehension between condition C4 (RX) and C5 (RIX) (\textit{p\textsubscript{comp,con}} = 0.98). Therefore, this hypothesis is not confirmed. 

\begin{table}[b]
\begin{tabularx}{\columnwidth}{XC{0.9cm}C{0.9cm}C{0.9cm}C{0.9cm}C{0.9cm}}
\textbf{Condition} & \textbf{Baseline} & \textbf{C1} & \textbf{C2} & \textbf{C3} & \textbf{C4} \\ \midrule
\textbf{Cronbach's $\alpha$} & 0.79 & 0.75 & 0.67 & 0.65 & 0.51 \\ 
\end{tabularx}
\caption{Cronbach's $\alpha$ for questionnaire measures across all conditions.}
\label{crotable} 
\end{table}

\subsubsection*{$H_{4}$: The correlation of self-report measures for textual explanation solutions will equally hold for testing the interactive solution.} We had to reject this hypothesis as well. Reacting to the large variance in C4 (cp. Figure~\ref{fig:rel-con}), we tested the validity of the questionnaire measures. As opposed to previous research~\cite{kizilcec_how_2016}, the self-reported measures are not correlated, instead they differ significantly (cp. Table~\ref{crotable}). This especially concerns C4, our redesign (RIX), where variance was very high (Cronbach's $\alpha = 0.51$).

% \begin{table}[t]
% \begin{tabularx}{\columnwidth}{XXXXX}
% \textit{\textbf{Factor}} & \multicolumn{1}{l}{\textit{\textbf{Comp.}}} & \multicolumn{1}{l}{\textit{\textbf{Fair.}}} & \multicolumn{1}{l}{\textit{\textbf{Acc.}}} & \multicolumn{1}{l}{\textit{\textbf{Trust}}} \\ \hline
% \textbf{Comp.} & - & 0.19 (0.13) & 0.15 (0.23) & \textbf{0.33 (0.01)} \\
% \textbf{Fair.} & 0.19 (0.13) & - & \textbf{0.40 (\textless{}0.01)} & \textbf{0.60 (\textless{}0.01)} \\
% \textbf{Acc.} & 0.15 (0.23) & \textbf{0.40 (\textless{}0.01)} & - & 0.16 (0.19) \\
% \textbf{Trust} & \textbf{0.33 (0.01)} & \textbf{0.60 (\textless{}0.01)} & 0.16 (0.19) & -
% \end{tabularx}
% \caption{Spearman correlation coefficients and p-values for C2-C4 for self-report measures: \textit{Comprehension} (5-point Likert), \textit{Fairness}, \textit{Accuracy} and \textit{Trust} (7-point Likert).}
% \label{corrtable} 
% \end{table}
\begin{table}[t]
\begin{tabularx}{\columnwidth}{XXXXX}
\textbf{Factor} & \textbf{Comp.} & \textbf{Fair.} & \textbf{Acc.} & \textbf{Trust} \\ \hline
\textbf{Comp.} & - & 0.19 & 0.15 & \textbf{0.33}$^*$ \\
\textbf{Fair.} & 0.19 & - & \textbf{0.40}$^*$ & \textbf{0.60}$^*$ \\
\textbf{Acc.} & 0.15  & \textbf{0.40}$^*$ & - & 0.16 \\
\textbf{Trust} & \textbf{0.33}$^*$ & \textbf{0.60}$^*$ & 0.16 & -
\end{tabularx}
\caption{Spearman correlation coefficients and p-values for C2-C4 for self-report measures \textit{Comp.=Comprehension}, \textit{Fair.=Fairness}, \textit{Acc.=Accuracy} and \textit{Trust} with $^*$ for $p < 0.05$.}
\label{corrtable} 
\end{table}

\begin{figure}[]
\begin{center}
\includegraphics[width=8cm]{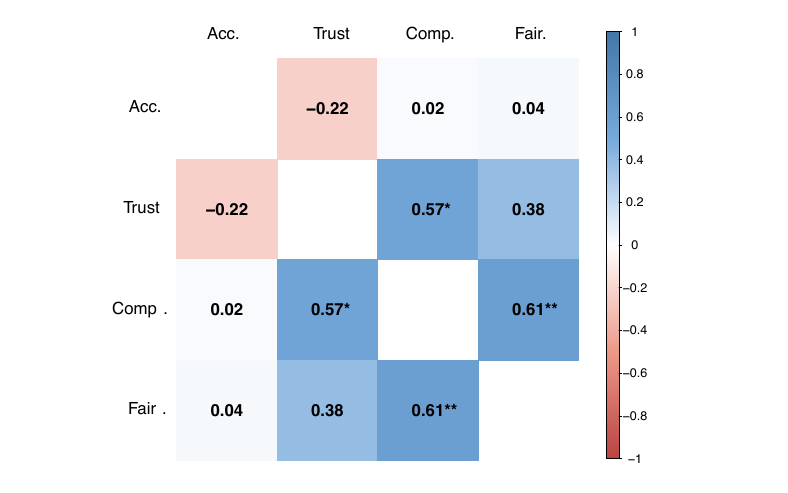}
\caption{Spearman correlation coefficient matrix for C4 with $^*$ for $p < 0.05$; $^{**}$ for $p < 0.005$.}
\label{fig:c4-corr}
\end{center}
\end{figure}

As a reaction to the high variance, we conducted Spearman's correlation tests for the ordinal Likert scales used in the participant self-report measures. 

The most statistically significant correlative finding from our data is that trust and fairness share a medium to strong positive relationship in our experiment. This is shown across all conditions (\textit{r\textsubscript{t,f} = 0.65, p < 0.01}), as well as between those wherein Recoin was introduced during onboarding (C2, C3, C4) (\textit{r\textsubscript{t,f} = 0.60, p < 0.01}) (cf. table \ref{corrtable}).

The predominant relationship of trust and fairness was reaffirmed for C2 and C3, the original design and the additional textual explanation respectively, as the strongest and most significant relationship (C2: \textit{r\textsubscript{t,f} = 0.73, p < 0.01}; C3: \textit{r\textsubscript{t,f} = 0.63, p < 0.01}).

A further relationship of fairness and accuracy was found for C2 and C3, as a medium positive relationship (C2: \textit{r\textsubscript{f,a} = 0.52, p = 0.01}; C3: \textit{r\textsubscript{f,a} = 0.53, p = 0.01}).

When participants used the interactive design of Recoin (RIX in C4), two different relationships emerged (cp. Figure~\ref{fig:c4-corr}). We found that the relationship between comprehension and fairness was strongest (\textit{r\textsubscript{c,f} = 0.61, p < 0.01}), closely followed by the relationship comprehension and trust (\textit{r\textsubscript{c,t} = 0.57, p = 0.01}). Surprisingly, the strong relationships found across all conditions were not present for the interactive design (cp. Figure~\ref{fig:c4-corr}).

%******************************************************************************
\subsubsection{Qualitative Analysis}
%\clmb{Jesse: Entweder, diese Kommentare werden noch mal richtig disktuiert oder ich würde sie rauslassen. Wenn Du sie drinlässt und noch mal näher diskutierst, musst noch sagen, wie Du sie ausgewählt hast. In der Form sind sie nichts halbes und nichts ganzes... Ich überlege auch ob sie nicht in eine Tabelle gehören, S1: Particpant Number, S2 Kommentare, S3... Bewertungen} 
Due to the high variance encountered in C4, and the unexpected lack of correlation between our self-report measures, we also expanded our analysis to participant statements. Accordingly, we sampled participant statements in order to probe specific subjective viewpoints. In this section, we showcase some preliminary insights.

\subsubsection*{Base Trust in Open Knowledge Platforms} A recurring theme, when participants chose to expand on their rating of trust, was a certain \textit{base trust} in open knowledge platform. This occurred even when no explanation element was provided, and also when participants received a poor grade in the condition that did not feature Recoin (Baseline):

\begin{quotation}
\emph{``Considering there was not a good definition of how we would be judged, it is tough to know if the judging was actually fair or unfair. However, I tend to trust Wikipedia so Wikidata is probably trustworthy.''} Baseline-P18; graded \textit{D}
\end{quotation}

This \textit{base trust} was also extended to the algorithm specifically, as long as it abides by platform standards:

\begin{quotation}
\emph{``I assume that an algorithm is used to grade the task, in which case I assume that it's free of bias, which is why I do trust Wikidata a good deal when it comes to fairness. Provided, the algorithm itself works as it's supposed to.''} C2-P15; rated \textit{Trust} at \textit{6 (High)}
\end{quotation}

\subsubsection*{High task efficiency may not indicate algorithm awareness} The qualitative data also suggests that task efficiency in terms of the algorithm does not necessarily indicate algorithm awareness. On the contrary, the only participant that offered a fundamentally accurate account of Recoin received the second lowest grade possible:

\begin{quotation}
\emph{``My only theory is that it's graded based on the relevance of entries made in regards to his occupation (astronaut) while most of my entries concerned his family, his awards and etc, rather than his activity as an astronaut.''} C2-P15; graded \textit{D}
\end{quotation}

The commentary of a well-performing participant (graded \textit{B}) furthermore suggests that there may be a difference in understanding algorithmic logic and understanding the integration into the algorithmic system:

\begin{quotation}
\emph{``It seems odd that I would be the one putting in the data and it is grading me considering why couldn't it just put the data in itself if it is accurate enough to grade.''} C1-P17; rated \textit{Accuracy} at \textit{6 (Very accurate)}
\end{quotation}

Finally, and in a similar fashion, a participant formulated the key question they had to the algorithmic system as follows:

\begin{quotation}
\emph{``I understand that the relevance is graded, I'm not sure exactly how relevance is judged.''} C2-P2; rated \textit{Comprehension} at \textit{2 (Low Understanding)}
\end{quotation}

In summary, the unexpectedly high variance in the C4 condition, combined with the difference in correlative relationships across conditions, as well as our qualitative data, allow us to gather relevant insights for further research. In the next section, we will discuss limitations to our experiment, before concluding with the contributions as well as the implications for future work.

%***************************************************************************************************************************************************
\section{Discussion}

First, we found no significant differences between the conditions in terms of average increases in completeness. However, this also suggests that the solution of textual explanation found in related work is not an inherently clear choice for algorithm awareness. This indicates that the design decisions for algorithm awareness are still methodologically unrefined.

Additionally, we sought to understand if our alternative to textual explanation, one taking an interactive and non-declarative approach, could be measured according to the existing self-report measures as suggested by previous research~\cite{kizilcec_how_2016}. We found that the measures ''Comprehension'', ''Fairness'', ''Accuracy'' and ''Trust'' were not equally distributed across our experimental conditions. On the contrary, divergent correlative relationships emerged. The status quo design (R1) as well as the addition of textual explanation design (RX) featured the same strong relationships of trust and fairness as well as fairness and accuracy. In contrast, our redesign (RIX) did not exhibit these relationships, but rather suggested that comprehension was most influential. This was shown in the medium to strong correlation between comprehension and fairness as well as comprehension and trust. We therefore posit that expanding on these self-report measures for algorithm awareness is another, distinct area requiring further research.

Moreover, the qualitative data we gathered also included insightful statements made by our participants. The phenomenon of \textit{base trust} that we encountered in participant statements is relevant for future algorithm awareness studies. If verified, it needs to be taken into account in cases where researchers may wish to abstract from platforms to look at specific problems.

In a broader context, experiments on transparency in algorithmic systems, especially in recommender systems, are frequently undertaken in order to minimize or even eliminate bias. However, as also found by Ekstrand and Tian in experimenting with various recommendation algorithms~\cite{ekstrand_all_2018}, a complete solution to the problem of bias is improbable. That is to say: bias is inevitable, and is a result of humans and technology interacting. This position is echoed in the work of the philosopher of technology Verbeek, who argues that technology fundamentally \textit{mediates} human relations to a particular ''world'', i.e., groups of other humans, values, practices etc.~\cite{verbeek_what_2006}. Biases, especially those commonly not aware of, play an instrumental role here. The solution, then, may not be finding the best measure for an elimination of bias, but rather finding the most actionable measure for making bias transparent. Our experimental results align with this assertion insofar that participants had issues with understanding the algorithmic system not on the basis of whether or not something is correctly calculated, but rather who or what has the \textit{agency} for judging the result (e.g., the platform itself, the algorithm as a contained unit, peer review etc.). This, along with a lack of significant differences between conditions, indicates that our intuition to design for an interactive mediation of the community-driven basis for Recoin was useful. Therefore, we posit that promoting algorithm awareness by interactivity is a promising research area.

Our study has a number of limitations that should be considered. %First, we used a relatively simple recommender system to reflect on the role of algorithm awareness. 
As opposed to other work (e.g. \cite{gunning_explainable_nodate}), our research focuses on non-technical experts. Furthermore, by recruiting our study participants over MTurk, it can be easily asserted that the demographics of the platform predispose the experiment to cultural bias. Additionally, online experiments in general are limited in two ways. On the one hand, observation of the subtleties of human-technology relations is not possible, such as the non-linguistic ways in which interaction expresses itself and decision-making occurs. On the other, by using MTurk we did not study Wikidata editors, but novices which might have never before come into contact with Wikidata. This means that, while we certainly could infer insights on algorithm awareness and human-technology relations, studying the lived practice of Wikidata editors may reveal other or even contradictory results. %Especially with regards to issues surrounding awareness, this endangers the validity of our results. However, given that we sought to establish a baseline on modes of algorithm representation in an online system, we considered this the best choice which might allow us to enroll the redesigned system on Wikidata. 

\section{Conclusion}
Our research was motivated by a wish to deepen our understanding of existing design parameters for algorithm awareness. We used the recommender system Recoin, employed in the online peer production system Wikidata, as a use case for our online experiments. In five different conditions, we provided the study participants with a varying degree of explanations and interactivity while using the recommender system. We were able to gather experimental data on the effect of various algorithm aware design measures, and to reflect on the validity of measures used in related work. However, our experiments alone are not yet exhaustive enough for us to reason more substantially about what human awareness means when algorithms are involved. 
Partly, this is due to the lack of longitudinal, qualitative data gathered from extensive and sustained use of Recoin. The participants of our experiments were predominantly unaware of Wikidata, and the task itself was both brief and controlled in terms of the knowledge that was provided. Wikidata lives and breathes from enthusiasts and domain experts that contribute extensively in their areas of interest. Thus, in future work, we seek to conduct studies that complement these results by probing individual and subjective use over time. This will allow us to understand more deeply, for example, how algorithm aware designs impact the relation between the Wikidata editors and the platform. From such studies, we plan to expand our framework to other use cases. Thereby, we hope to contribute to the urgent need for understanding how the increasingly ubiquitous algorithmic systems shape everyday life for and from the Web. 

\bibliographystyle{ACM-Reference-Format}
\bibliography{bibliography}

\end{document}